# Strain Mapping by Digital laser Speckle Correlation, Validation and Comparison

**Mahshad Mosayebi**

[1] Department of Mechanical Engineering and Energy Processes, Southern Illinois University
1230 Lincoln Drive
Carbondale, IL 62901
Tel. 618-453-7049; Email: mahshad.mosayebi@siu.edu

**ABSTRACT**

This Paper introduces a new Non-Contact, Optical method for displacement measurements, and strain mapping as well as comparing it to traditional Digital Image correlation (DIC) and laser interferometry measurement method. This Method incorporates diffracted laser speckle images from the surfaces through DIC to track displacement and locate strain values.
In order to evaluate the feasibility of the method, various experiments were done and results were compared to laser interferometry based and traditional DIC.
All the experiments were designed and done with affordable equipment, while they resulted in displacement as small as 30 μm detection.
Results presented by this paper are showing that DilSIC is an economic, accurate, rapid and applicable method for the mentioned purpose. Since it does not require any artifact speckle pattern, it can be used on the non- accessible area, limits and difficulties for creating speckle pattern is not applicable for this technique.
In this research, various magnitude of strains have been examined within the range of [0-10%]. As this technique is a hybrid method of DIC and Laser speckle measurement it could eliminate some of the limits that everyone has. Those removed restrictions include but not limited to being able to measure strain within range of [0-10%] while using fringes on laser speckle does not let the measurement exceed 2%. Also using laser speckle pattern can end all the challenges to achieve the qualified speckle pattern as they can be adjusted to match the requirement easily [1].

## INTRODUCTION

Exceeding the ultimate strain in loading is one of the most common reasons for a part failure. Therefore, developing an in situe method for strain mapping can lead to prevent the future damages and failure.
Equation (1) is the general equation used for strain measurement under the assumption of small displacement where the Euler and Lagrangian approaches match.
Therelation between displacements $u_k$ and strains $\varepsilon_{kl}$ with k, l = x, y, z takes the form

$$\varepsilon_{kl} = 1/2\,(u_{k,l} + u_{l,k})$$

Equation (1), Strain and displacement

As what this equation discloses, tracking displacement is the main step for strain mapping. While Lagrangian refrence frame is offering contact methods such as strain gage, Euler derivetive of the strain equation can be used in non-contact method such as DIC (add the DIC equation), Speckle Interferometry, Photo- Elasticity, X-ray diffraction and Holographic Interferometry [2].

Digital Image correlation is a measurement technique has been used and known worldwide for displacement and strain mapping [], material property recognitions and defect detection. Not being limited to homogenous and isotropic material, being full-field method can be named as of the advantages of DIC. DIC compares images captured from the object surface before and after applying load. This comparison leads to track deformation and strain measurement on the surface [3-9]. Traditional DIC required fabricated speckle pattern as tools for tracking. Though using this tool is an acceptable and validated technique for DIC, it brought some restrictions. Sometimes we do not have access to the part to apply artifact speckle pattern, or on many cases applying material is not allowed [bio ref]. Also achieving qualified speckle pattern is challenging [10,11]. More of all due to the size of the speckle pattern small deformation is hard to track.

In order to remove some of the restrictions mentioned above, non-contact measurement tools were investigated.

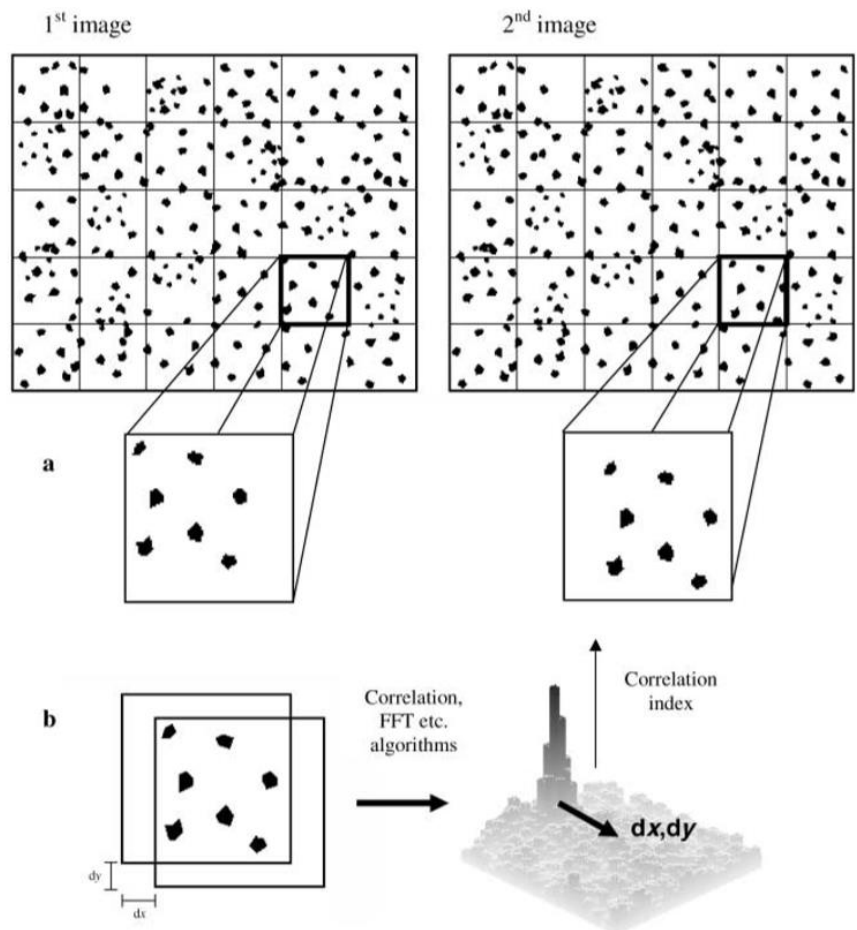

Figure [1] (a) Image correlation is a displacement mapping technique. To create a displacement map, the two images obtained at different strains are divided into smaller subregions. (b) Pairs of subregions are then compared computationally, using correlation or FFT algorithms. The displacement vector joins the center of the subregion and the point of highest correlation. This operation is repeated for all subregion pairs to create a displacement map [13]

As can be seen in figure 1, in traditional DIC, speckle patterns are fabricated to the specimen. The process of applying artifact speckle pattern causes some limits and restrictions [14] which led us to find an alternative way to create speckle patter.

On the other hand, Laser interferometry has been used widely for strain measurement. These laser-based methods are well known for the accuracy and ability to find the small strain values. Combination of these methods results in developing a hybrid method that uses laser speckle pattern up to track deformation through DIC the following sections are disclosing the validation and comparison steps for this new technique.

Meanwhile, laser-based methods are also used to find defects. This method uses laser speckle interferometry of two laser beams to find the deformation [15] [16]. This method is popular for its accuracy and ability to find the nano- scale deformation. Laser based techniques are widely used due to their high accuracy, feasibility and sensitivity; however, using fringes to measure strain has an important drawback: it will not be accurate in strain values higher than 2% [17].

Individual limitations and similarities of the two methods led to the development of the new hybrid technique which could improve the laser based and DIC techniques a great deal. This hybrid technique takes advantage of laser speckles to track deformation, while the cost-effective setup is being used. [18-21]].

## EXPERIMENTAL DETAIL

In order to validate the method performance and precision, four executive experiments have been done. Results are compared to laser interferometry and DIC Analysis in the similar conditions [20]. Translation test, tensile test and strain concentration test as well as defect detection test. Figure [2] is showing the set up for translation test. (pull up the exact detail from the thesis).

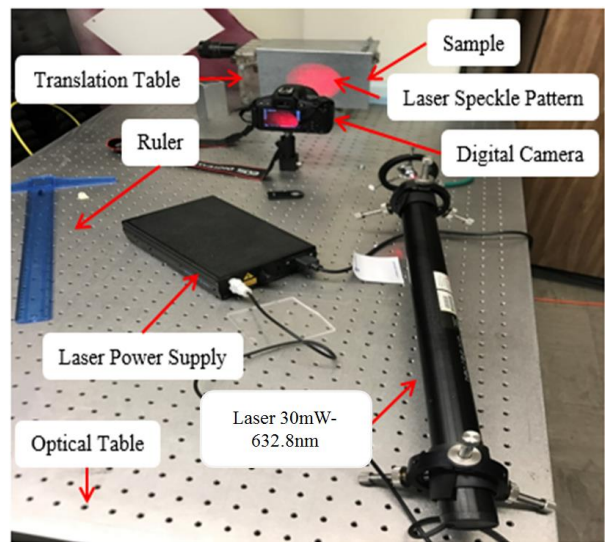

**Figure 2: General Setup for Tensile Test, 40X expanding lens, 632.8nm-30 mW laser, T5I Cannon camera- Lens is EFS 55mm, observation beam, (camera) must be perpendicular to the target surface [23,24]**

## Translation Evaluation

In translation test, system was expected to track displacement of the part on the translation table with 2.54-micron accuracy. To show the method performance capability in tracking displacement for different surface roughness, variable materials (and finishes) have been used such as Al sheet, uniformly painted Al sheet and rubber. After validating the performance of DiLSIC in translation detection, we moved forward with rubber material (due to high flexibility) with in the rest of steps (find from the thesis). Displacement was applied in the domain of [30micron and 25,400 micron) to evaluate the system limits. The sample geometry was a regular 126mm × 25.4mm rectangle. Results of the translation test can be found in figure 2.

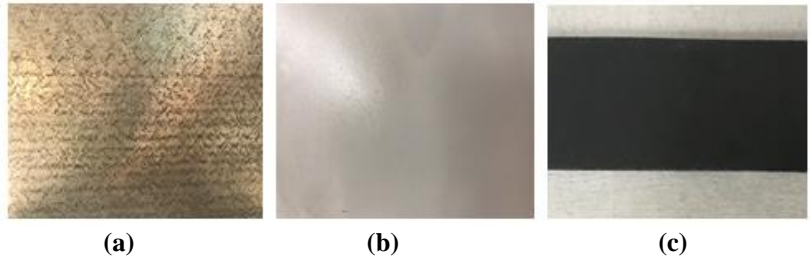

(a)          (b)          (c)

**Figure 3: Sample used in translation test (a) Aluminum sheet with texture (b) Uniform painted aluminum sheet (c) Rubber sample on the white background**

The camera used in all parts was T5I Cannon camera- Lens is EFS 55mm. To find the best setup for the camera, images with different apparatus size and exposure time were captured. Then, the resulting histograms showed the best setting for the camera setup as 8 for aperture size and 1.3 seconds for time of exposure.

Below figures show the results for translation evaluation through DiLSIC (both numerical and visual). Matching results with the actual displacement are a proof of DiLSIC capability in tracking displacement.
While theoretical calculations proof that incorporating images through DIC cannot exceed 0.02-pixel size, only 0.4% error in the results can be considered as a great accuracy and capability.

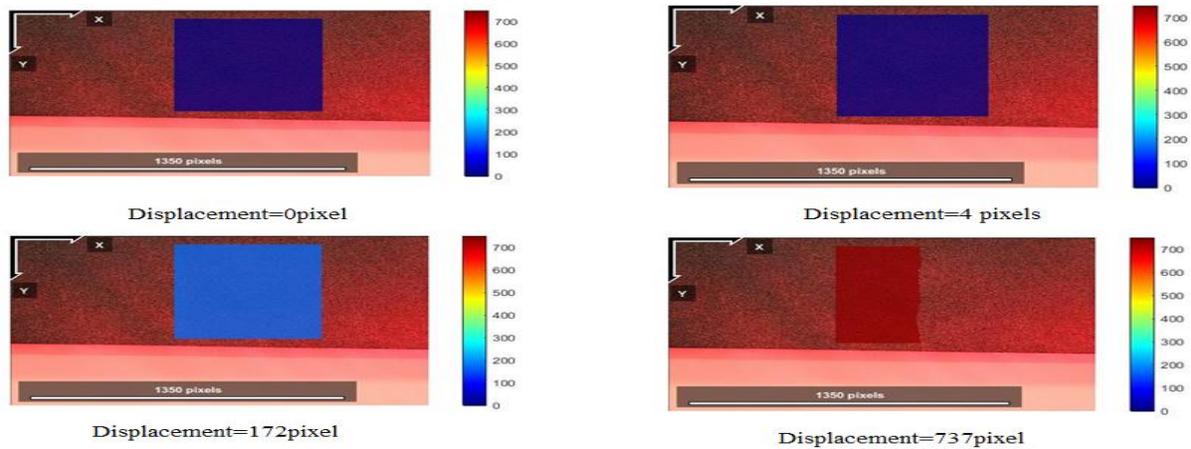

Figure 4: Resulting displacement contours for rubber sample in translation test. Pixel size was 0.031mm. The uniform color of contours shows the rigid body motion.

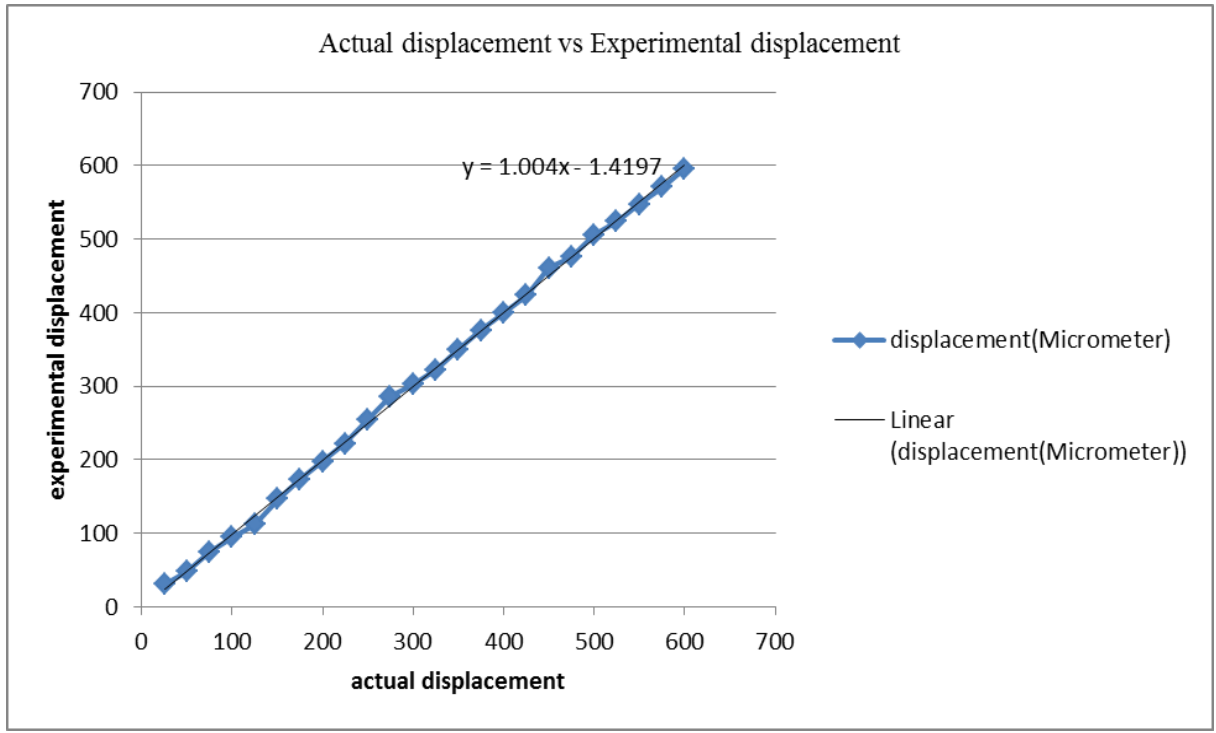

**Figure 5: Comparison between resulted displacement with DiLSIC and actual displacement chart**

One of the main limits of using traditional laser interferometry is not being able to recognize translation movement. This fact restricts inspected specimen to be fixed on a support. While in many cases such as in vitro bone analysis, it cannot be guaranteed [22].

## *Tensile Test*

Tensile test has been done with rubber to evaluate the method performance in strain measurement. Results were compared to DIC for validation. Displacement load was applied along the X axis to the one end of the sample while the other end was fixed at the tensile test device support. The device accuracy is 25.5 micron as well. Strains of 1% to 10% were applied to one end of the sample. The same tests were done with the traditional DIC. Figure [6] is showing the DIC and DiLSIC samples. As can be seen in this figure, paint has been applied to the DIC sample as the fabricated Speckle Pattern

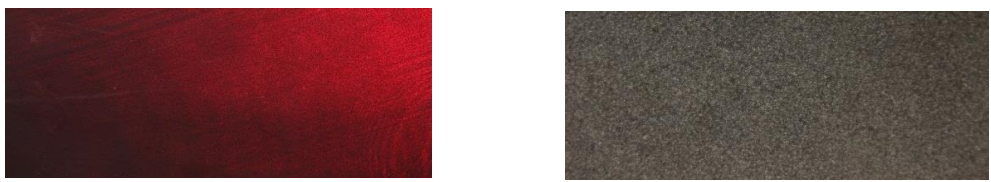

**Figure 6: Sample used in tensile test with DIC (right) and sample used for DilSIC (left)**

Figures below show the visual and numerical comparison between DiLSIC and DIC in strain measurement. As can be found, DiLSIC could achieve more comparative results in strain less than 2% in compare to traditional DIC. They both achieve good results in strain more than 2% while laser interferometry is not capable of measuring strain more than 2% [16].

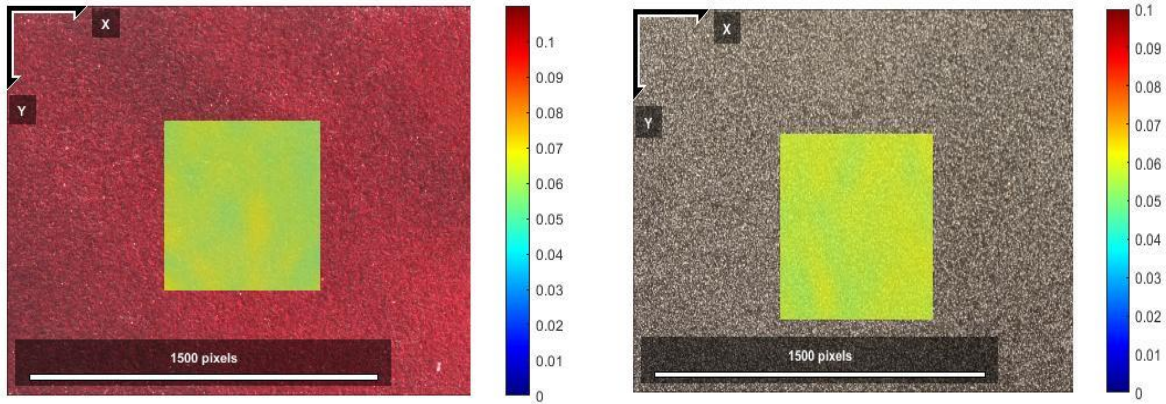

**Figure 7: Comparison between DIC (right) and DiLSIC (left) resulting contours after tension. Actual strain in both cases was 6%**

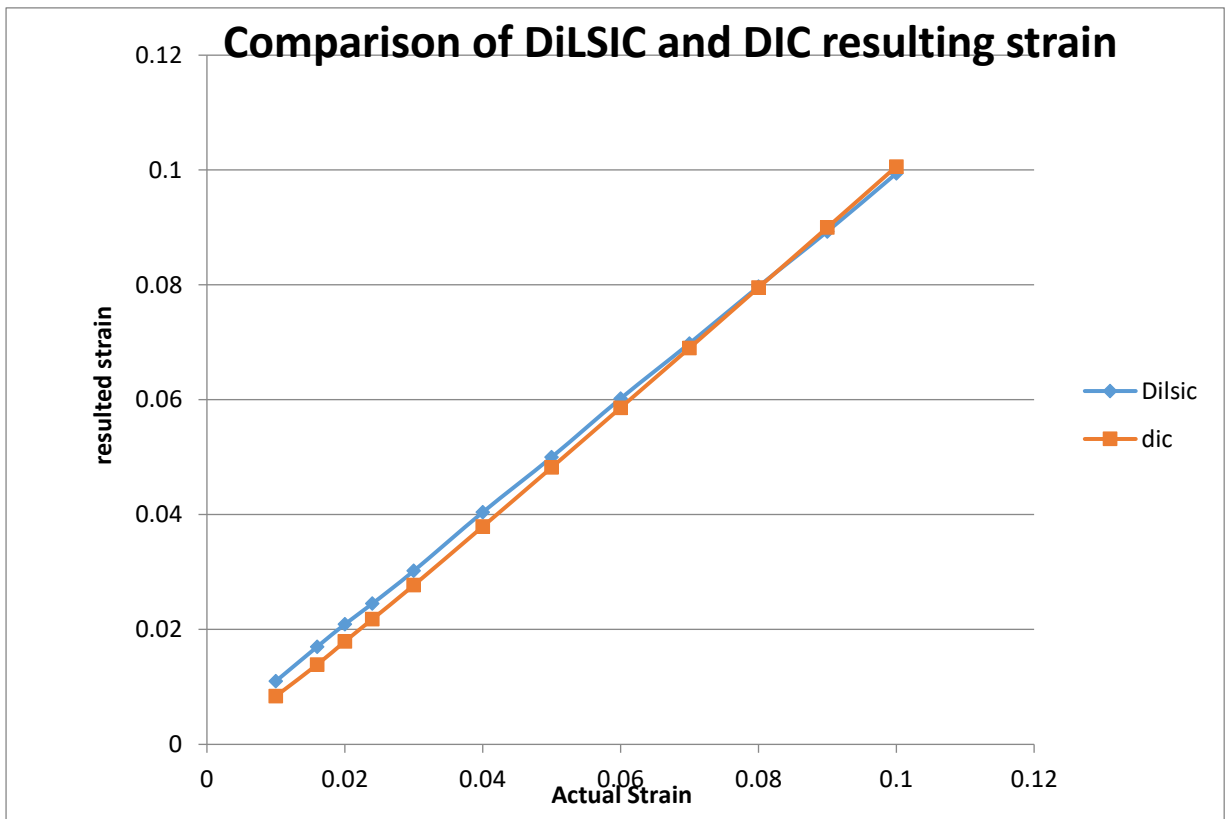

**Figure 8: Strain comparison between DiLSIC and DIC chart**

It is also worth mentioning that in tensile test for strain measurement, DiLSIC resulted in values with mean error of 1% while the same value for traditional DIC was 1.8%. [20]

## Strain Concentration

In this step, a 10mm diameter hole was created on the samples to investigate DiLSIC performance in strain mapping. It was expected to be able to detect the strain concentration as well as locating the critical area (points). Again, results were compared to DIC results. Random points aligned an imaginary vertical line passing from the hole center were picked for the strain value comparison with the points at the same coordinates at the DIC sample

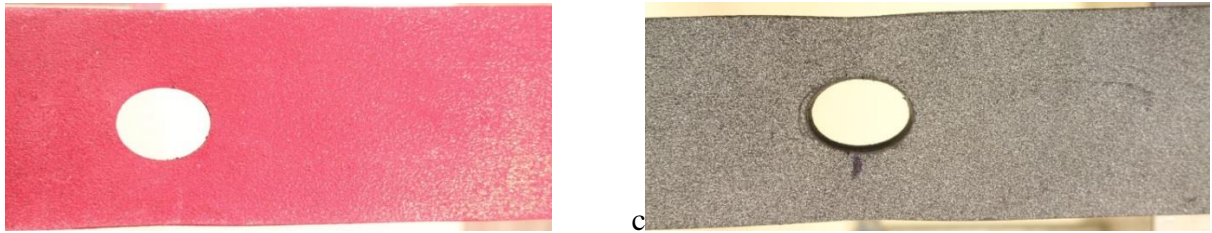

**Figure 9: Deformed sample in tensile test in DiLSIC (left) and DIC (right) method.**

Figure [10] and [11] are presenting the results and comparison of strain mapping and strain concentration detection in both DIC and DiLSIC methode. High agreement in visual contours and numerical values is a further proof and validation for DiLSIC capability in strain mapping.

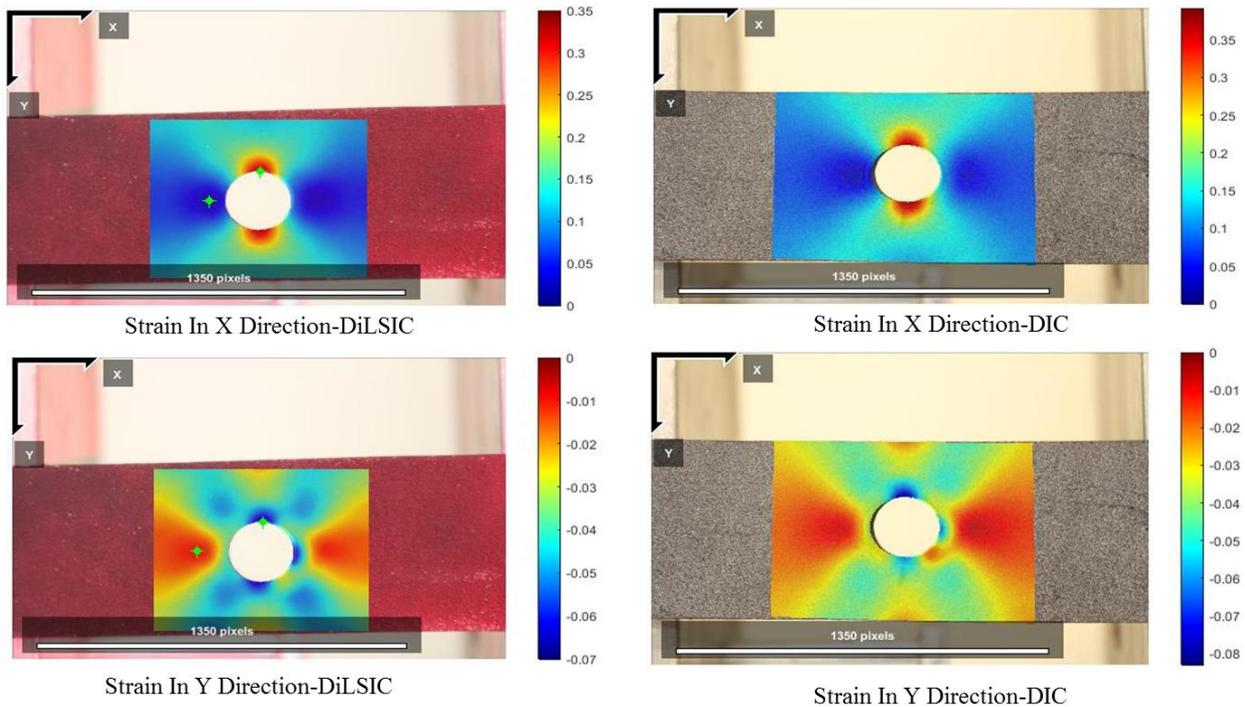

Strain In X Direction-DiLSIC    Strain In X Direction-DIC

Strain In Y Direction-DiLSIC    Strain In Y Direction-DIC

**Figure 10: Strain concentration contour- DIC (left) and DiLSIC (right) results comparison**

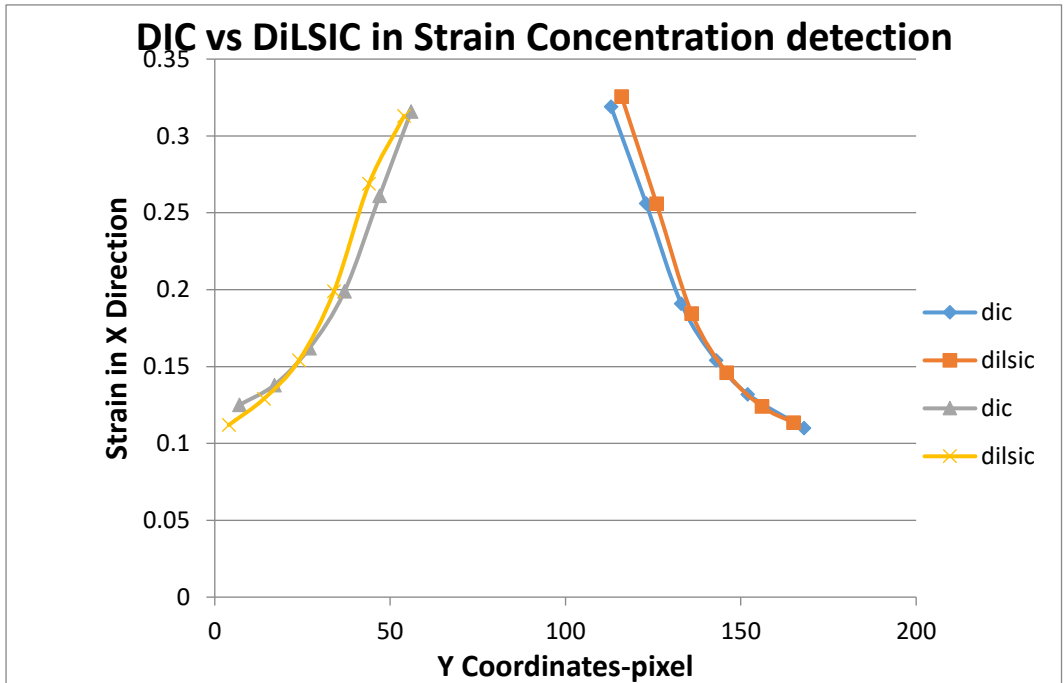

**Figure 11:** Comparison between DIC and DiLSIC in strain concentration detection. The high agreement between results is the other evidence of DiLSIC performance precision.
.

## CONCLUSION

A new hybrid method was evaluated and compared to a few other strain mapping techniques. This method could combine the advantages of both laser based and DIC methods to eliminate some of the problems from each individual method. The performance of this technique in finding defects, strain measurement, and strain concentration was evaluated in all experiments. The comparison between DiLSIC and DIC and traditional laser interferometry measurement methods is presented in the table below.

| criteria | Traditional Laser Interferometry | DIC | DiLSIC |
|---|---|---|---|
| Capable of translation recognition | No | Yes | Yes |
| Specimen preparation | Not Required | Required | Not Required |
| Capability of Strain Measurement Range | Only less than 2% | [0%-10%] Has been validated | Only more than 2% |
| Strain Measurement Accuracy | Maximum for Values less than 2% | 1.8% error | 1% error |

## REFERENCES

r

[1]. JinlianSong[a], JianhongYang[a], FujiaLiu[a], KefeiLu[c], "Quality assessment of laser speckle patterns for digital image correlation by a Multi-Factor Fusion Index", Optics and Lasers in Engineering Volume 124, January 2020, 105822

[2]. H. Weisbecker· B. Cazzolato· S. Wildy· S. Marburg·J. Codrington· A. Kotousov "Surface Strain Measurements Using a 3D Scanning Laser Vibrometer, 6 October 2011 © Society for Experimental Mechanics 2011

[3]. J. Perie, S. Calloch, C. Cluzel and F. Hild, 2002," *Analysis of a Multiaxial Test on a C/C Composite by Using Digital Image Correlation and a Damage Model"*, Vol 42, No 3. Universite Paris 6, Cedex, France.

[4]. F. Laurin, J.-s. Charrier, D. Leveque, J.-F. Mair, A. Mavel, P. Nunez, 2012, " *Determination of the properties of composite materials thanks to digital image correlation measurements*", Procedia IUTAM 4 , Cedex, France

[5]. J. Tyson and T. Schmidt, *Optical Deformation & Strain Measurement in Biomechanics*, GOM mbH

[6]. T. C. Chu, W. F. Ranson, M. A. Sutton and W. H. Peters, "*APPLICATIONS OF DIGITAL-IMAGE-CORRELATION TECHNIQUES TO EXPERIMENTAL MECHANICS*," Experimental Mechanics, vol. 3, no. 25, pp. 232-244, 1985.

[7]. T. Chu, A. Mahajan and C. T. Liu, "*An Economical Vision-Based Method to Obtain Whole-Field Deformation Profiles*,," Experimental Techniques, vol. 6, no. 26, pp. 25-28, 2002.

[8]. J .L. W. Carter, Michael. D. Uchic, Michael J. Mills," *Impact of Speckle Pattern Parameters on DIC Strain Resolution Calculated from in-situ SEM experiments*", The Ohio State University, Columbos, OH

[9]. T. A. Berfield, J. K. Patel, R. G. Shimmin and etc, 2007, "*Micro and Nanoscale measurement of Surface and Internal Plane via Digital Image Correlation*", Society for experimental Mechanics

[10]. J. QUINTA DA FONSECA, P. M. MUMMERY & P. J. WITHERS," Full-field strain mapping by optical correlation of micrographs acquired during deformation", Journal of Microscopy, Vol. 218, Pt 1 April 2005

[11]. Y. Su, X. Xu and Q. Zhang , 2016"*Quality assessment of speckle patterns for DIC by consideration of both systematic errors and random errors*", Optics and Lasers in Engineering, University of science and Technology, China

[12]. H. Taheri, F. Delfanian and J. Du, 2013, "*Wireless NDI for Aircraft Inspection",* ASNT 22[nd] research symposium

[13]. E. Archbold & A.E. Ennos (1972) Displacement Measurement from Double exposure Laser Photographs, Optica Acta: International Journal of Optics, 19:4, 253-271, DOI: 10.1080/713818559

[14]. J. QUINTA DA FONSECA, P. M. MUMMERY & P. J. WITHERS " Full-field strain mapping by optical correlation of micrographs acquired during deformation" Journal of Microscopy, Vol. 218, Pt 1 April 2005

[15].

[16]. S. Yoshida, Muchair, I. Muhamad, R. Widiastuti and A. Kusnowo, " *Optical Interferometric technique for deformation Analysis*", Research and development center for applied physics, Serpong, Indonasia

[17]. J. Gryagoridis, 212, "*Laser Based Nondestructive Inspection Technique*" , Springer Science+ Business Media, LLC, Department of Mechanical Engineering, Cape Peninsula University of Technology, Cape Town, South Africa

[18]. J. Brillaud and F. Lagattu, 2002, "*Limits and Possibilities of Laser Speckle and White-Light image correlation methods: Theory and experiments*", Optical Society of America

[19]. L. M. Richards, S. M. Shams Kazmi, J. L. Davis , A. K. Dunn and K. E. Olin, 2010, "*Low-cost Laser Speckle Contrast Imaging of blood flow using a webcam*", Department of Biomedical Engineering, The University of Texas at Austin, Austin, TX
[20]. W. Dandach , J. Molimard and P. Picart, ," Direct Strain And Slope Measurement Using 3D DSPSI", LTDS, UMR 5513, École Nationale Supérieure des Mines, SMS-EMSE, CNRS, Saint-Étienne, France
[21]. P. L. Reu, and B. D. Hansche, "*Digital Image Correlation combined with Electronic Speckle Pattern interferometry for 3D Deformation Measurement in Small Samples*", Sandia National Laboratories, PO Box 5800, Albuquerque, NM 87185
[22]. Y. Y. Hung "Shearography: a new optical method for strain measurement and nondestructive testing" ,School of Engineering Oakland University Rochester, Michigan 48063

[23]. Mahshad Mosayebi, Seyed Fouad Karimian, and Tsuchin Philip Chu, "NDT Using Digital Laser Speckle Image Correlation (DiLSIC)", ASNT 26[th] Research Symposium, March 2017
[24]. M. Mosayebi, "Digital Laser Speckle Image Correlation", Southern Illinois University